\documentclass[prl,aps,twocolumn]{revtex4}
\usepackage{ifthen}
\usepackage{graphicx}

\begin{document}
\title{Traces of stimulated bosonic exciton-scattering in semiconductor luminescence}
\author{D.~H\"{a}gele}
\affiliation{Institut f\"{u}r Festk\"{o}rperphysik,
Universit\"{a}t Hannover, Appelstra\ss e 2, D-30167 Hannover,
Germany}
\author{S. Pfalz}
\affiliation{Institut f\"{u}r Festk\"{o}rperphysik,
Universit\"{a}t Hannover, Appelstra\ss e 2, D-30167 Hannover,
Germany}
\author{M.~Oestreich}
\affiliation{Institut f\"{u}r Festk\"{o}rperphysik,
Universit\"{a}t Hannover, Appelstra\ss e 2, D-30167 Hannover,
Germany}
\begin{abstract}
We observe signatures of stimulated bosonic scattering of
excitons,  a precursor of Bose-Einstein-Condensation (BEC), in the
photoluminescence of semiconductor quantum wells. The optical
decay of a spinless molecule of two excitons (biexciton) into an
exciton and a photon with opposite angular momenta is subject to
bosonic enhancement in the presence of other excitons. In a spin
polarized gas of excitons the bosonic enhancement breaks the
symmetry of two equivalent decay channels leading to circularly
polarized luminescence of the biexciton with the sign opposite to
the excitonic luminescence. Comparison of experiment and many body
theory proves stimulated scattering of excitons, but excludes the
presence of a fully condensed BEC-like state.
\end{abstract}
\maketitle
In the last few decades intensive theoretical and experimental
work has been devoted to the understanding and observation of a
macroscopic ground state of correlated electrons and holes in
semiconductor structures. Analogous to Bose-Einstein-Condensation
(BEC) of atoms, a dilute gas of bound electron-hole pairs - so
called excitons - is predicted to condense into a common ground
state \cite{griffin95,hanamuraPR77,zimmermannPSS70}. Observation
of condensation had been reported in CuO$_2$, where excitons with
extraordinary long radiative lifetime exhibit signatures of BEC
\cite{snokePRL90}. More recently, the observation of a
macroscopically ordered state in a GaAs based heterostructures was
interpreted in terms of exciton condensation
\cite{butovNATURE03_2}. However, all reports on possible exciton
condensation are controversially discussed and alternative
interpretations of the experimental data have been given
\cite{oharaPRB00,butovPRL04}. Clear criteria that are able to give
unambiguous evidence for bosonic effects and ultimately BEC are
therefore highly desirable. A precursor of excitonic BEC is
stimulated scattering of excitons into other exciton states. The
scattering rate is proportional to a bosonic enhancement factor
$(1+n)$, where $n$ is the number of excitons in the final state.
In this letter we present experimental evidence for stimulated
bosonic scattering in the exciton system of optically excited
semiconductor quantum wells (QWs). The optical decay of a spinless
biexciton (molecule of two excitons) into an exciton and a photon
with opposite spin is subject to bosonic enhancement. The presence
of spin polarized excitons causes the biexciton to decay
preferentially into an exciton with the same spin. The resulting
imbalance of circularly polarized photons in the final state
leaves as signature a finite degree of circular polarization at
the biexciton photoluminescence (PL) line. We report on clear
experimental evidence for the described effect in semiconductor
QWs. A full quantum mechanical many body model yields a criterion
for bosonic effects which by comparison with our data proofs
stimulated scattering but excludes the presence of a BEC-like
state.

We investigate stimulated bosonic scattering in the
photoluminescene of a high quality 10 nm thick ZnSe QW embedded in
500 nm ZnS$_{0.07}$Se$_{0.93}$ barriers grown by molecular beam
epitaxy (MBE) on GaAs substrate \cite{breuningPRB02}. The sample
is kept in a finger cryostat and excited with frequency doubled fs
pulses from an 80 MHz Ti:Sapphire laser through a microscope
objective to obtain high exciton densities. The spectral width of
the pulses is narrowed down to below 4~meV by means of a pulse
shaper to allow for resonant excitation of bound electrons and
holes close to the exciton resonance which avoids the creation of
hot carriers with excess energy. The polarization of the exciting
light is carefully controlled by a Soleil-Babinet retarder. The
semiconductor luminescence is measured in reflection geometry by a
synchroscan streakcamera providing a spectral and temporal
resolution of 2~meV and 6~ps, respectively. By passing the PL
through an electrically tunable liquid-crystal retarder, the left
and right circularly polarized components are separately measured
without artifacts from possible beam displacement known from
traditional optical elements like a $\lambda/4$-wave-plate.

\begin{figure}
   \centering
   \includegraphics[width=8.5cm]{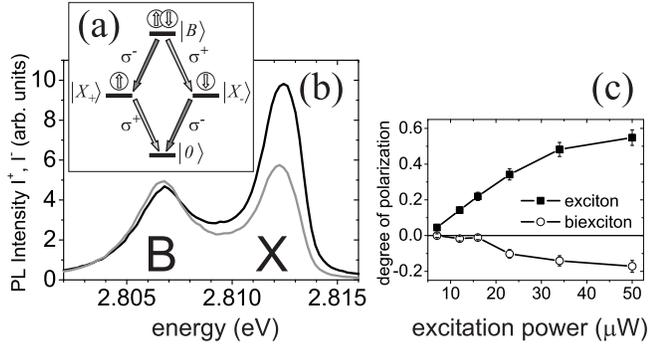}
   \caption{ a) optical decay scheme, b) polarization resolved PL spectrum at T =10 K,
    20~ps after laser excitation. Black line: $\sigma^+$ polarization. Gray line: $\sigma^-$ polarization.
    c) density dependence of
    polarization 15~ps after excitation, exciton (full squares) and biexciton (open circles). }
    \label{fig1}
\end{figure}

We optically excite excitons with angular momentum $J_z = 1$ by
pumping the sample with  circularly polarized laser pulses of
about 2~ps duration. Excitons with $J_z = -1$ are also created due
to imperfect selection rules and spin relaxation. Subsequently
biexcitons are formed by binding of excitons with opposite angular
momentum. Excitons with parallel angular momenta cannot form a
bound state. The spin related criteria for binding are analogous
to hydrogen where only atoms with antiparallel electron spin form
molecules. Figure 1(b) depicts a PL spectrum 20 ps after
excitation at a sample temperature of 10~K. The high energy
luminescence peak results from the optical recombination of
excitons. Because of the small photon momentum only excitons with
momentum $K\approx0$ contribute to the emission. The low energy
luminescence peak results from the decay of a biexciton into a
photon and a free exciton. The energetic position of the peak is
at $E_x - \delta_{\rm bi}$, where $E_x$ is the exciton energy and
$\delta_{\rm bi}$ is the biexciton binding energy. Because the
biexciton consists of two excitons with opposite angular momenta,
one decay channel results in an emitted $\sigma^+$ polarized
photon and a remaining $J_z = -1$ exciton, whereas the other
channel yields the opposite angular momenta ($\sigma^-$ and $J_z =
1$). Within a four level scheme, the average optical emission of
biexcitons should be completely unpolarized [Fig. \ref{fig1}(a)].

In contrast, we find for high excitation densities {\it circular
polarized emission} at the biexciton line, whose sign is {\it
opposite} to that of the excitons. Figure \ref{fig1}(b) shows the
circular polarized PL spectrum 20~ps after quasi resonant
excitation  of the exciton resonance with 50~$\mu$W creating an
estimated exciton density of about $8 \times 10^{11}$~cm$^{-2}$.
The time delay guarantees that the initial coherence of laser
pulse and excitons has completely vanished. The spectrum clearly
shows the polarization effect.
The magnitude of biexciton
polarization increases with increasing exciton density from zero
to about 17~\% [Fig. \ref{fig1}(c)]. The simultaneous increase of
polarization at the exciton can be explained by the increasing
fraction of excitons with opposite spin that are bound into
biexcitons and therefore do not contribute to excitonic emission
with opposite circular polarization. For low densities, the degree
of exciton polarization is much lower than expected from the
selection rules. We explain this by a short electron spin
relaxation time $\tau_s$ which decreases from about 100~ps to
20~ps at lowest densities \cite{relaxationnote}. Independently, we
observed the same biexciton polarization effect also in 5~nm
ZnSe/ZnMgSSe multiple QWs and in a 9.9~nm GaAs/AlAs single QW.

 In the following we
exclude a number of artifacts that might have caused the polarized
emission at the biexciton line. (i) Creating excitons with
opposite circular polarized laser pulses changes both, the sign of
exciton and biexciton photoluminescence. This exclude that the
biexciton polarization is caused by some polarizing element in the
optical path of the luminescence or by some asymmetry which could
have favored negative polarization. Similarly, the application of
an in-plane magnetic field leads to spin-oscillations which
modulate both the exciton and biexciton polarization with the same
frequency \cite{heberlePRL94}. (ii) The optical dielectric
function of a QW is after excitation with circular polarized light
no longer the same for $\sigma^+$ and $\sigma^-$ polarized light.
Since the rate of spontaneous emission depends on refractive index
and reabsorption, we calculated emission at the biexcitonic line
taking into account the polarization dependent dielectric function
and the dielectric properties of the sample structure
\cite{strouckenPRB96}. We found a low degree of polarization
induced by the spin dependent sample properties, that, however,
exhibited the same sign as the exciton polarization. I.e., the
surprising polarization effect of the biexciton may be even larger
than it appears from the bare luminescence data. (iii) A large
energetic spin splitting of the exciton levels induced by exchange
interaction would lead to energetically distinct fully polarized
biexciton lines with equal intensity, but no average polarization.
A finite splitting could therefore explain polarized emission in
the low energy tail of the biexciton line. This artifact is
excluded, because, first, there is no visible splitting of the
$\sigma^\pm$ exciton lines, and, second, integration over the full
biexciton line still exhibits polarized emission.

We will first give a preliminary explanation of polarized emission
at the biexciton line based on the Dicke theory of spontaneous
emission, which originally was developed to treat cooperative
optical decay of several multi-level atoms in a small volume
\cite{dickePR54,agarwalOC76,mandelwolf95}. Consider a chain of $N$
adjacent sites which can be occupied by an exciton with either
spin 1, spin -1, or a biexciton \cite{darkexcitonnote}. The chain
may be viewed as a quantum wire with the length of $N$ exciton
units and periodic boundary conditions. In this picture the decay
of an exciton at a certain site corresponds to the transition from
an excited atomic level $|X_{j,\pm}\rangle$ to the ground state
$|0_i\rangle$. The PL intensity of the full exciton chain at the
exciton line is given by $I_{\pm}=\langle R_{\pm}^\dagger R_{\pm}
\rangle$, where $R_{\pm} = \sum_{i=1}^N |0_i\rangle\langle
X_{i,\pm}|$
is the macroscopic dipole operator of all $N$ excitons and $i$
labels the exciton sites. Photon emission at the biexciton line is
linked to the annihilation of a biexciton, ${X}_{j,+} {X}_{j,-}$,
and the creation of an exciton, ${X}^\dagger_{j,+}$, and a photon
with opposite angular momentum. Within the Dicke formalism one
needs to calculate $I_{{\rm b},\pm} = \langle R_{{\rm
b},\pm}^\dagger R_{{\rm b},\pm} \rangle$, where the macroscopic
dipole operator $R_{{\rm b} ,\pm} = \sum_{j=1}
{X}_{j,\mp}^{\dagger}{X}_{j,+} {X}_{j,-} $ is now expressed in
second quantization with the operator relations $X_{i,\pm}
X_{j,\pm} = X_{j,\pm} X_{i,\pm}$ and $X_{i,\pm}^2 = 0$ (hard core
bosons). Exciton and biexciton operators can be defined in
momentum space $x_{k,\pm}  = N^{-1/2}\sum_{j=1}^N {X}_{j,\pm}
\exp\left(2\pi i k
  j/N\right)$, $b_{k}  =  N^{-1/2}\sum_{j=1}^N {X}_{j,+} {X}_{j,-}\exp \left(2\pi i k
  j/N\right)$ with $k=0,...,\pm (N-1)/2$,
which is closer to the usual notation in semiconductor theory.
In this representation we find for biexciton luminescence $I_{{\rm
b},\pm} = \sum_{k,k'} \langle b^\dagger_k x_{k,\mp}
x_{k',\mp}^\dagger b_{k'} \rangle $, implying momentum
conservation as expected. For calculating PL of a sample with
incoherent excitons and biexcitons we can drop off-diagonal terms
with $k \neq k'$. We proceed calculating the $k$-dependent terms $
I_{k,\pm} = \langle b^\dagger_k x_{k,\mp} x_{k,\mp}^\dagger b_{k}
\rangle $ by applying the exact Bose-commutation relations
$x_{k,\mp} x^{\dagger}_{k,\mp} - x^{\dagger}_{k,\mp} x_{k,\mp} = 1
- 2 n_\mp/N$, where the number operator $n_\mp = \sum_k
x_{k,\mp}^\dagger x_{k,\mp}$ counts all excitons with the same
spin (including excitons bound in biexcitons). After approximating
the four operator expectation values by factorized products, the
biexciton PL intensity obtains the final form
\begin{eqnarray}
 I_{k,\pm} & = & \langle
b^\dagger_k (1+x_{k,\mp}^\dagger
x_{k,\mp}- 2 n_\mp /N) b_{k} \rangle \nonumber \\
& \approx & \langle b^\dagger_k b_{k} \rangle \left( 1+\langle
x_{k,\mp}^\dagger x_{k,\mp} \rangle  - 2 \frac{\langle n_\mp
\rangle}{N}\right). \label{theory}
\end{eqnarray}
It follows that biexciton emission $I_{k,\pm}$ depends not only on
the occupation of biexcitons $\langle b^\dagger_k b_{k} \rangle$
but also on a enhancement factor $\left( 1+\langle
x_{k,\mp}^\dagger x_{k,\mp} \rangle - 2 \langle n_\mp \rangle / N
\right)$. We interpret this factor as stimulated bosonic
scattering of final state excitons with momentum $k$ into excitons
that were already present in the state $k$. This immediately
explains the negative circular polarized biexciton PL in the
presence of spin polarized excitons with positive angular momenta
$\langle x_{k,+}^\dagger x_{k,+} \rangle
> \langle x_{k,-}^\dagger x_{k,-} \rangle$. Similarly, positive circular polarization is observed
in the presence of excitons with negative angular momenta.  Only
for highest exciton densities [$\langle n_\mp \rangle/N \approx
50$\%] the third term of the enhancement factor may dominate and
lead in special cases to polarized biexciton emission with the
same sign as the excitons. Our observations exhibit a clear
deviation of the usual proportionality between occupation number
and PL intensity in semiconductors (without a laser-like optical
feedback).

The derivation above already gives a good impression of the origin
of the polarization effect. However, the calculation does not
regard exciton energy dispersion, and makes no prediction about
the actual exciton-biexciton distribution which at high densities
is not simply described by a Bose-Einstein distribution. Excitons
are no ideal bosons and the Pauli exclusion principle applies to
their fermionic constituents, electrons and holes. Inclusion of
such features is unfortunately beyond the applicability of full
microscopic theories of semiconductor luminescence that neglect
spin and correlations related to biexcitons
\cite{kiraPQE99,hannewaldPRB00,fernandezSSC98,olayacastroPRL01}.
We therefore introduce a simplified model Hamiltonian for a 1D
exciton biexciton system
\begin{equation}
 H  =  \sum_{k,\sigma=\pm}  E_k x^\dagger_{k,\sigma} x_{k,\sigma}
- \sum_{j} \delta_{\rm b} X_{j,+}^\dagger X_{j,+}\,
X_{j,-}^\dagger
 X_{j,-},
\end{equation}
that quantitatively includes the exciton dispersion $E_k$.
Biexciton formation is included by a short range (on-site)
potential well with depth $-\delta_{\rm b}$, which in principle
allows for a biexciton state that extends in contrast to the Dicke
model over more than just one site.
The distance between two adjacent sites is assumed to be four
exciton Bohr radii $4 a_0$, which correspond to kinetic exciton
momenta of $p_{k} = \pi \hbar k /2 a_0 N$ \cite{noteSpacing}. We
obtain with the experimental values for exciton radius $a_0 =
5$~nm and the exciton mass $m_X = 0.344 m_0$ the exciton
dispersion relation $E_k = E_{\rm X} + p_k^2 / 2 m_X$ with a
maximum kinetic energy of about $E_{\rm max} = 11$~meV and $E_{\rm
X} = 2.8$~eV. In order to calculate the luminescence spectrum of
the system at a given temperature, we calculate the multi particle
density matrix
\begin{equation}
 \rho \propto \exp\left(-\frac{1}{k_{\rm B} T}\left(H - \mu_+ n_+ -
 \mu_-
 n_-\right)\right)
\end{equation}
where $\mu_+$ and $\mu_-$ are the chemical potentials which govern
the average number of spin up $\langle n_+ \rangle$ and spin down
$\langle n_- \rangle$ excitons.
The intensity of $\sigma^\pm$ polarized PL emitted by decay of a
multi-particle energy eigenstate $|\psi_i\rangle$ into a state
$|\psi_f\rangle$ is given by $I^\pm_{i,f} = \left|\langle
\psi_f|D_\pm|\psi_i\rangle\right|^2,$ where $D_\pm = d \sum
X_{j,\pm}$ is the overall dipole operator (in rotating wave
approximation). Consequently, the full polarized luminescence
spectrum that belongs to the mixed state $\rho$ is given by
\begin{equation}
I_\pm(\epsilon) = \sum_{i,f} \langle \psi_i|\rho|\psi_i\rangle
\left|\langle \psi_f|D_\pm|\psi_i\rangle\right|^2
\lambda\left(\epsilon-(\epsilon_i - \epsilon_f)\right),
\end{equation}
where $\lambda$ has the homogenous linewidth of an optical
emission event and $\epsilon_i$ is the energy of the multi-exciton
states.
\begin{figure}
   \centering
   \includegraphics[width=7cm]{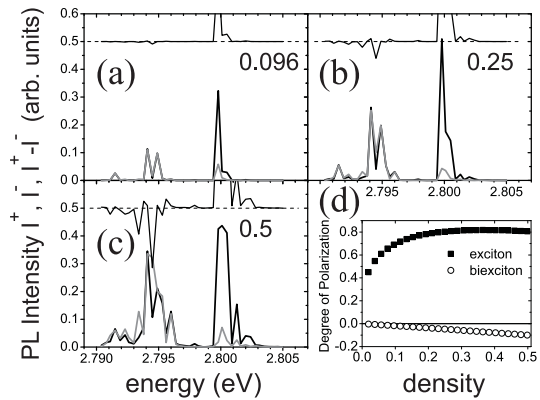}
   \caption{a)-c) Calculated polarized PL spectra $I^+$, $I^-$, and $I^+ - I^-$
   (offset by 0.5 for better visibility) at T = 15 K for increasing densities
   of 0.096, 0.25, and 0.5 excitons per site. d)
   density dependence of the excitonic and biexcitonic degree of polarization.}
    \label{FigPol25T15}
\end{figure}
Figure \ref{FigPol25T15} shows the calculated PL spectra for
$\sigma^+$ and $\sigma^-$ polarization for the model with $N=5$
sites ($2^{2\cdot 5} \times 2^{2\cdot 5} $ density matrix) at a
temperature of $T=15$~K and with a fixed spin polarization
$(\langle n^+ \rangle - \langle n^- \rangle )/(\langle n^+ \rangle
+ \langle n^- \rangle ) = 0.25$. At low densities $\langle
x_{k,\mp}^\dagger x_{k,\mp} \rangle \ll 1$ there is almost no
biexciton polarization effect as expected [compare eq.
(\ref{theory})]. At elevated densities, the polarization effect
increases significantly to up to 10~\% at a density of $0.5$
excitons per site (2.5 excitons distribute over all 5 sites),
which corresponds to about $1.25 \times 10^{12}$~cm$^{-2}$.  The
calculated magnitude of the biexciton polarization effect in the
different density regimes compares qualitatively well with
experiment. Also the relative height of exciton and biexciton peak
compares well with our measurement.
 We obtain improved
agreement with data if we assume an exciton temperature of $T =
30$~K in the calculations, which can be understood by the increase
of carrier temperature during the binding process of biexcitons.

\begin{figure}
   \centering
   \includegraphics[width=7cm]{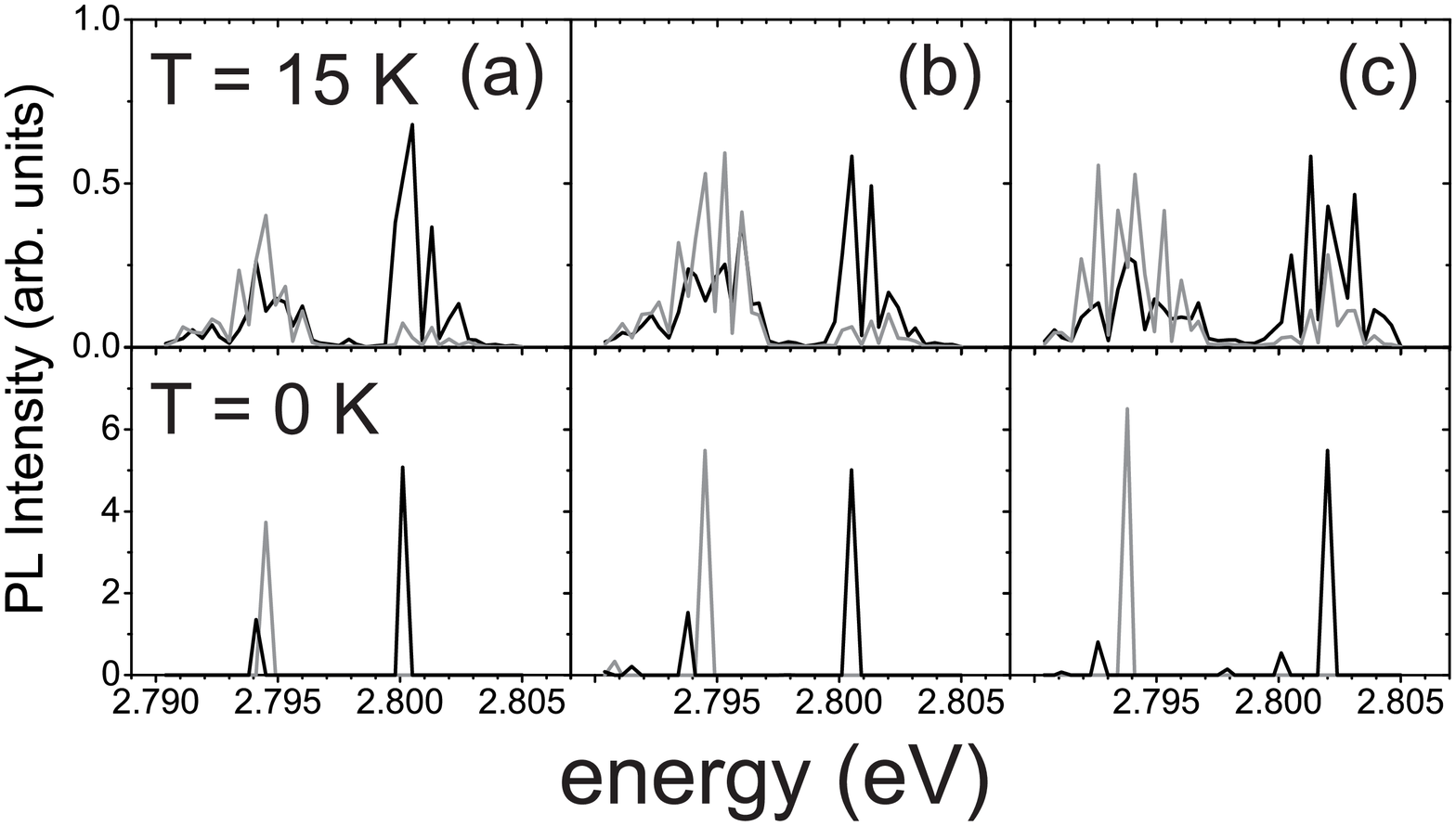}
   \caption{Calculated PL spectra at T = 15 and T = 0 K. a) Number of excitons ($n_+$, $n_-$) = (2,1)
   b) (3,2) c) (4,2).}
    \label{fig15vs0}
\end{figure}

Next we discuss criteria to observe BEC-like condensation in the
exciton biexciton system. First we show that the calculated ground
state of our model Hamilton can be well approximated by an
analytical expression which has the from of a phase average
BCS-like state (BCS state becomes a BEC state in the low density
limit)
\begin{equation}
|\psi(\phi_1,\phi_2)\rangle \propto \prod_{i=1}^N \left(1+c_1 e^{i
\phi_1} X_{i,+}^\dagger + c_2 e^{i \phi_2} X_{i,+}^\dagger
X_{i,-}^\dagger \right)|0\rangle,
\end{equation}
which contains excitons and biexcitons as condensing entities with
two macroscopic phases $\phi_1$ and $\phi_2$.
Since the Hamilton $H$ commutes with the particle operators
$n_\pm$, the ground state is required to posses sharp particle
numbers which we obtain by phase-averaging
$|\psi(\phi_1,\phi_2)\rangle$
\begin{equation}
 |\Phi(n,m)\rangle \propto \int_{\phi_1,\phi_2=0}^{2 \pi}
  \hspace{-1cm} |\psi(\phi_1,\phi_2)\rangle e^{ -i (n-m) \phi_1 - i
  m \phi_2 } d\phi_1\,d\phi_2,\label{bcsstate}
\end{equation}
where $n$ and $m$ are the number of excitons with spin $1$ and
spin $-1$, respectively. The overlap $\alpha = |\langle
\Phi(m,n)|\chi(m,n)\rangle|^2$ of the BCS-like state with the
numerically gained ground states $|\chi(m,n)\rangle$ exhibits good
agreement ($\alpha$ ranging from 0.6 and 1.0). The calculated
emission spectra for $|\chi(m,n)\rangle$ with $n > m$ reveal at
$T=0$ sharp peaks at the exciton and biexciton energy [fig.
\ref{fig15vs0}, bottom row].
 The biexciton peak shows a high
degree of polarization which increases with density. The presence
of strongly polarized biexciton emission is a clear signature for
a highly correlated multi exciton state. The exciton peak is 100
\% polarized, because all minority excitons with opposite spin
form a biexciton at zero temperature. For comparison, a system
with the same average exciton densities, but $T = 15$~K exhibits
broad spectral features and a much lower degree of polarization at
the biexciton and exciton resonance [fig. \ref{fig15vs0}, top
row]. Consequently, we suggest as a criterion for a condensed
system of partially polarized exciton the observation of sharp PL
lines with a high degree of polarization at the exciton {\it and}
biexciton emission PL line. This criterion is experimentally less
demanding than the suggested measurement of the photon statistics
of emitted light \cite{fernandezSSC98,olayacastroPRL01}. Applying
the criterion to our data, we conclude that we observe stimulated
bosonic scattering of excitons, but do not observe a fully
condensed multi-exciton state.

In conclusion we found unambiguous traces of stimulated bosonic
scattering of excitons in the photoluminescence of a
photo-generated exciton biexciton system. The spectral signature
at the biexciton line is explained in terms of cooperatively
radiating excitons and biexcitons within Dicke theory and
numerically treated within a many body theory which yields all
major features of our experimental observations.

We gratefully acknowledge discussions with M. Baranov, Raymond
Chiao, and A. Knorr, as well as support by W. W. R\"uhle and H.
Kalt in the early stage of this work. The work is supported by the
DFG (German Science Foundation) and the BMBF.


%
%
%

\begin{thebibliography}{10}
\bibitem{relaxationnote}
Long spin lifetimes at high density might be due to a motional
narrowing effect
  caused by increasing exciton scattering.
\bibitem{darkexcitonnote}
Excitons with $\pm2$ spin are not taken into account as they do
not contribute
  to PL.
\bibitem{noteSpacing}
The exciton spacing is somewhat arbitrary. A larger spacing
corresponds to a
  broader potential well for biexciton binding and a lower maximum density of
  excitons that can be described in our model.
\end{thebibliography}


\end{document}